\documentclass[pre,aps,twocolumn,showpacs,floatfix,10pt]{revtex4-1}
\usepackage{graphicx}
\usepackage{color}
\usepackage{amsmath, amsthm, amssymb}
\newcommand{\be}{\begin{equation}}
\newcommand{\ee}{\end{equation}}
\newcommand{\bea}{\begin{eqnarray}}
\newcommand{\eea}{\end{eqnarray}}

\begin{document}

\title{Re-entrant Disordered Phase in a  System of Repulsive Rods
on a Bethe-like Lattice}
\author{Joyjit Kundu}
\email{joyjit@imsc.res.in}
\affiliation{The Institute of Mathematical Sciences, C.I.T. Campus,
Taramani, Chennai 600113, India}
\author{R. Rajesh}
\email{rrajesh@imsc.res.in}
\affiliation{The Institute of Mathematical Sciences, C.I.T. Campus,
Taramani, Chennai 600113, India}

\date{\today}

\begin{abstract}
We solve exactly a model of monodispersed rigid rods of length $k$
with repulsive interactions on the random locally tree like layered 
lattice. For $k\geq 4$ 
we show that with increasing 
density, the system undergoes two phase transitions: first from a low density
disordered phase to an intermediate density nematic phase and second from
the nematic phase to a high density re-entrant
disordered phase. When the coordination number is $4$, both the phase 
transitions are continuous
and in the mean field Ising universality class. For even coordination number
larger than $4$, the first transition is discontinuous  while the nature of the
second transition depends on the rod length $k$ and the interaction parameters.
\end{abstract}


\maketitle

\section{Introduction}

A system of long hard rods in three dimensions undergoes a phase transition 
from a disordered phase with no orientational order 
to an orientationally ordered nematic phase as the density of rods is
increased beyond a critical
value~\cite{onsager1949,flory1956a,zwanzig1963}, and has applications in
the theory of liquid crystals~\cite{vroege1992,degennesBook}.
In two dimensions, though an ordered phase that breaks a continuous 
symmetry is disallowed~\cite{mermin1966}, the system
undergoes a Kosterlitz-Thouless type transition from an isotropic phase with 
exponential 
decay of orientational correlation to a high density critical 
phase~\cite{straley1971,frenkel1985,khandkar2005,vink2009}. 
On two-dimensional  lattices, remarkably, there are two entropy driven transitions for
long rods:
first from a low density disordered (LDD) phase to an intermediate 
density nematic phase, and second  from the nematic phase 
to a high density disordered (HDD)
phase~\cite{ghosh2007}. While the existence of the first transition
has been proved rigorously~\cite{giuliani2012}, the second transition
has been demonstrated only numerically~\cite{joyjit2013}.
In this paper, we consider a model of rods interacting via a repulsive 
potential on the random
locally tree like layered lattice, and through an exact solution show
the existence of two phase transitions as the density is varied.

We describe the lattice problem in more detail. 
Rods occupying $k$ consecutive lattice sites along any lattice direction will 
be called  $k$-mers. No two $k$-mers are allowed to intersect, and all
allowed configurations have the same energy. For dimers 
($k$ = 2), it is known that the system remains disordered at all
packing densities~\cite{lieb1972}. 
For $k \geq k_{min}$, it was argued that the
system of hard rods would undergo two phase transitions as density is
increased~\cite{ghosh2007}. On both the square and the triangular 
lattices $k_{min}=7$~\cite{ghosh2007,fernandez2008c}. 
Monte Carlo studies show that the first transition from LDD
phase to nematic phase is continuous, and is in the 
Ising universality class for the square lattice and in the three-state Potts model
universality class for the triangular
lattice~\cite{fernandez2008a,fernandez2008b,fernandez2008c,linares2008,fischer2009}.
The existence of this transition has been 
has been proved rigorously for large $k$~\cite{giuliani2012}.
The second transition from nematic to HDD phase was studied using an
efficient algorithm that ensures equilibration of
the system at  densities  close to full packing~\cite{joyjit_dae,joyjit2013}. 
On the square lattice the second 
transition is continuous with effective critical exponents that are
different from the two dimensional Ising exponents, 
though a crossover to the Ising universality class
at larger length scales could not be ruled out~\cite{joyjit2013}. 
On the triangular lattice the second transition 
is continuous and the  critical exponents are numerically close to 
those of the first transition. This
raises the question whether the LDD and HDD phases are same
or different. 

Is there a solvable model of $k$-mers that shows two
transitions with increasing density and throws light on the HDD phase? 
The hard core $k$-mer problem was solved exactly on the random locally
tree like layered lattice (RLTL), a Bethe-like
lattice~\cite{rajesh2011}. This lattice was introduced because 
a uniform nematic order is unstable on the more 
conventional Bethe lattice when the coordination number is 
larger than $4$. 
However, on the RLTL, while a stable
nematic phase exists for all even coordination numbers greater than or equal to four,
the second transition is absent for hard rods~\cite{rajesh2011}. 
In this paper, we relax the hard-core
constraint and allow $k$-mers of different orientations to
intersect  at a lattice site. Weights $u, v, \ldots$ are associated with 
sites that
are occupied by two, three, $\ldots$ $k$-mers.  When the weights are zero, we
recover the hard rod problem. We solve this
model on the RLTL and show that for a range of $u, v, \ldots$, the system
undergoes two transitions  as the density is increased:
first from a LDD phase to
a nematic phase and second from the nematic phase to a HDD phase.
For coordination number $q=4$, the two transitions are continuous and
belong to the mean field Ising universality class. 
For $q\geq 6$, where $q$ is an even integer,
while the first transition is first order, the  second transition is
first order or continuous depending on the values of $k, u, v, \ldots.$.  
In all cases, it is possible to continuously transform  the LDD phase
into the HDD phase in the
$\rho$--interaction parameters phase diagram without crossing any
phase boundary, showing that the LDD and HDD phases are qualitatively
similar, and hence the HDD phase is a re-entrant LDD phase.

The rest of the paper is organized as follows.
In Sec.~\ref{sec:model}, we recapitulate the construction of RLTL and
formulate the model of rods on this lattice. In Sec ~\ref{sec:q_4}, we 
derive the analytic expression for free energy for fixed density of
horizontal and vertical $k$-mers on the $4$-coordinated RLTL.
It is shown that the system undergoes two continuous
phase transitions for $k\geq 4$. In
Sec.~\ref{sec:q_6}, the free energy is 
computed for
coordination number $q=6$, 
and the dependence of the
nature of the transition on the different parameters are detailed.
Sec.~\ref{sec:summary} summarizes 
the main results of the paper, and discusses some possible extensions. 
  
\section{\label{sec:model} The RLTL and 
definition of the model}

\begin{figure}
\includegraphics[width=\linewidth]{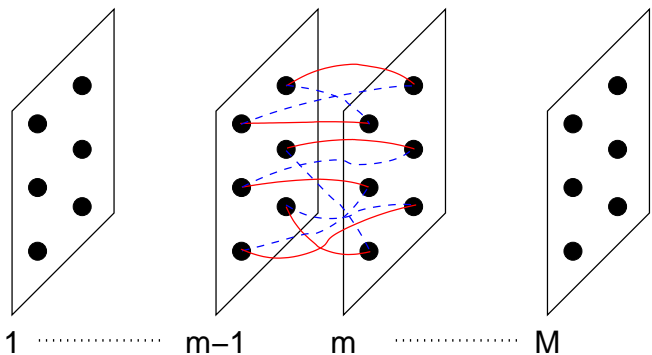} 
\caption{\label{fig:rltl}Schematic diagram of the
RLTL with $N=6$ sites per layer 
and coordination number $4$. A typical bond configuration between layers 
$m-1$ and $m$ is shown with $X$ bonds in red (solid) lines and $Y$
bonds in blue (dotted) lines.
}
\end{figure}

The RLTL was introduced in Ref.~\cite{rajesh2011}. In this section,
we recapitulate its construction for coordination number $q=4$. Generalization 
to larger even values of $q$ is straightforward. Consider a collection of $M$ 
layers, each having $N$ sites. A layer $m$ is connected to its adjacent 
layer ($m-1$) by $N$ bonds of type $X$ and $N$ bonds of type $Y$. Each site 
in the $m^{th}$ layer is connected with exactly one randomly chosen site in the 
$(m-1)^{th}$ layer with a bond of type $X$. Similarly bonds of type $Y$ are also connected 
by random pairing of sites in the two adjacent layers. Hence, the total number 
of such possible pairing between two layers is $(N!)^2$. 
A typical bond configuration is
shown in Fig.~\ref{fig:rltl}. For a $q$-coordinated lattice with
periodic boundary conditions, the total 
number of different possible graphs is $(N!)^{q M/2}$, and with open boundary 
conditions there are $(N!)^{q(M-1)/2}$ different possible graphs. In the 
thermodynamic limit, the RLTL contains few short loops 
and locally resembles a Bethe lattice. 

We consider a system of monodispersed rods of length $k$ on 
the RLTL. A $k$-mer 
occupies $(k-1)$ consecutive bonds of same type.
Rods on $X$ ($Y$) type of bonds will be called $x$-mers ($y$-mers).
Weights $e^{\mu_1}$ and $e^{\mu_2}$ are associated with   
each $x$-mer and $y$-mer, where $\mu$'s are chemical potentials.
Linear rods 
comprising of $k$ monomers are placed on the RLTL such that a site can be
occupied by utmost two $k$-mers.
Two $k$-mers of the same type can not intersect.  
A weight $u$ is associated with every site that is occupied 
by two $k$-mers of different type.  
The limiting case $u=0$ corresponds to the hard core problem. 
For even $q \geq 6$, a site can be occupied by utmost $q/2$ $k$-mers,
each
of different type.

Consider the annealed model on RLTL.  
The average partition function is
\be
Z_{av}(M,N) = \frac{1}{N_{\mathcal{R}}}\sum_{\mathcal{R}}
Z_{\mathcal{R}}(M,N),
\ee
where $Z_{\mathcal{R}}(M,N)$ is the partition function for a given
bond configuration $\mathcal{R}$ and $N_{\mathcal{R}}$ 
is the number of different bond configurations on
the lattice.  In the
thermodynamic limit the mean free energy per site is obtained by
\be
f= - \lim_{M,N\to\infty}\frac{1}{MN}\ln Z_{av},
\label{eq:free_defn}
\ee
where the temperature and Boltzmann constant have been set equal to $1$.

\section{\label{sec:q_4}$k$-mers on RLTL with coordination number $4$}

In this section, we calculate the free energy of the system on the 
RLTL of coordination number $4$ for fixed $u$ and fixed densities of $x$-mers and
$y$-mers. The phase diagram of the system is obtained by minimizing the
free energy with respect to $x$-mer and $y$-mer densities for a fixed total density.

\subsection{\label{subsec:free energy_q_4}Calculation of Free energy}

To calculate the partition function, consider the operation of adding the 
$m^{th}$ layer, given the configuration up to the $(m-1)^{th}$ layer. The 
number of ways of adding the $m^{th}$ layer is denoted by $C_m$. $C_m$
will be a function of the number of $x$-mers and $y$-mers passing
through the $m^{th}$ layer and the number of intersections between
$x$-mers and $y$-mers at the $m^{th}$ layer.

Let $x_m$ $(y_m)$ be the number of $x$-mers ($y$-mers) whose 
left most sites or heads are in the 
$m^{th}$ layer. $X_m$ and $Y_m$ are the number of sites in the $m^{th}$ layer 
occupied by $x$-mers and $y$-mers respectively, but where the site is not the 
head of the $k$-mer.  Clearly,
\be
X_m = \sum_{j=1}^{k-1}x_{m-j},~ 
Y_m = \sum_{j=1}^{k-1}y_{m-j},~ 1\leq m \leq M,
\ee
with $x_m=y_m=0$, for $m\leq 0$. To have all $k$-mer fully contained with 
in the lattice for open boundary condition we need to impose, $x_m=y_m=0$ for, 
$m\geq M-k+2$. 

In a $k$-mer, let $h$ denote its head or  left most site 
and $b$ denote the other $k-1$
sites. Then, we define $\Gamma^m_{ij}$, where $i,j=h,b$, to be the
number of intersections at the $m^{th}$ layer between site $i$ of an
$x$-mer and site $j$ of a $y$-mer. For instance, 
$\Gamma^m_{hh}$ is the number of sites in the $m^{th}$ layer, occupied
simultaneously by the heads of an $x$-mer and  a $y$-mer.

Given $\{x_m\}$, $\{y_m\}$ and $\{\Gamma^m_{ij}\}$, the calculation of
$C_m$ reduces to an enumeration problem. The details of the
enumeration are given in appendix~\ref{sec:appendix1}. We obtain
\begin{widetext}
\bea
C_m& =& \frac{N! X_m! Y_m! (N-X_m)! (N-Y_m)!} 
{
(x_m-\Gamma^m_{hh}-\Gamma^m_{hb})!
(y_m-\Gamma^m_{hh}-\Gamma^m_{bh})! (X_m-\Gamma^m_{bb}-\Gamma^m_{bh})!
(Y_m-\Gamma^m_{bb}-\Gamma^m_{hb})!
} \nonumber\\
&& \times \frac{1}{\left(N-X_m-Y_m-x_m-y_m+
\displaystyle\sum_{i,j=b,h} \Gamma^m_{ij}\right)!
\displaystyle\prod_{i,j=b,h}\Gamma^m_{ij}!
}.
\label{eq:cm}
\eea
\end{widetext}
The partition function is then the weighted sum of the product of
$C_m$ for different layers:
\bea
Z_{av}& = & \frac{1}{(N!)^{2M}} 
\sum_{\{x_m\}, \{y_m\}, \{\Gamma^m_{ij}\}} \nonumber \\
&&\prod_{m} \left( C_m e^{\mu_1 x_m} 
e^{\mu_2 y_m} u^{\sum _{ij}\Gamma^m_{ij}} \right).
\label{eq:partition}
\eea
where the sum is over all possible number of $x$-mers, $y$-mers and
number of doubly occupied sites.
Since the summand is of order $\exp(NM)$, for large $N, M$, we 
replace the summation with the largest summand with negligible error. 
For the summand to be maximum 
with respect to $x_l$, we set:
\be
\frac{C(\{x_m+\delta_{m,l}\},\{y_m\},\{\Gamma^m_{ij}\}) e^{\mu_1}}
{C(\{x_m\}, \{y_m\},\{\Gamma^m_{ij}\})} \approx 1,
\label{eq:maximize_c}
\ee
where $C=\prod_m C_m$.
Likewise, we can write equations for each of the variables.

We look for homogeneous solutions such that $\rho_x=x_m k/N$, $\rho_y=
y_m k/N $, and $\gamma^m_{ij}= \Gamma_{ij}/N$ are variables
that are
independent of $N$ and have no spatial dependence. Here $\rho_x$ and
$\rho_y$ are fractions of sites in any layer that are occupied by
$x$-mers and $y$-mers respectively. In terms of these 
variables, Eq.~(\ref{eq:maximize_c}) and the corresponding one for
$y_j$ reduce to
\be
\frac{(\rho_x-\frac{\rho_x}{k})^{k-1}(1-\rho\!+\!\sum_{ij} \!\!\gamma_{ij})^k
(\frac{\rho_x}{k}\!-\!\gamma_{hh}\!-\!\gamma_{hb})^{-1}
}
{(1-\rho_x+\frac{\rho_x}{k})^{k-1}
(\rho_x-\frac{\rho_x}{k}-\gamma_{bb}-\gamma_{bh})^{k-1}
}
= e^{-\mu_1},
\label{eq:fixed_x}
\ee
and
\be
\frac{(\rho_y- \frac{\rho_y}{k})^{k-1}(1- \rho \!+ \!
\sum_{ij} \!\! \gamma_{ij})^k
(\frac{\rho_y}{k}\!-\!\gamma_{hh}\!-\!\gamma_{bh})^{-1}
}
{(1-\rho_y+ \frac{\rho_y}{k})^{k-1}
(\rho_y- \frac{\rho_y}{k}-\gamma_{bb}-\gamma_{hb})^{k-1}}=
e^{-\mu_2},
\label{eq:fixed_y}
\ee
where $\rho=\rho_x+\rho_y$ is the total density. On maximizing
the summand in Eq.~(\ref{eq:partition}) with respect to $\Gamma^l_{ij}$, we obtain
\begin{subequations}
\label{eq:gamma}
\bea
\frac{
[\rho_x (1\!\!-\!\frac{1}{k}\!)\!-\!\gamma_{bb}\!-\!\!\gamma_{bh}]
[\rho_y (1\!\!-\!\frac{1}{k}\!)\!-\!\gamma_{bb}\!-\!\!\gamma_{hb}]
}
{ \gamma_{bb}(1-\rho+\sum_{ij}\gamma_{ij}) } 
&=&\frac{1}{u}, \label{eq:fixed_bb}\\
\frac{(\frac{\rho_x}{k}-\gamma_{hh}-\gamma_{hb})(\frac{\rho_y}{k}-\gamma_{hh}-
\gamma_{bh})}
{ \gamma_{hh}(1-\rho+\sum_{ij}\gamma_{ij}) } 
&=&\frac{1}{u}, \label{eq:fixed_hh}\\
\frac{(\frac{\rho_x}{k}-\gamma_{hh}-\gamma_{hb})[\rho_y(1-\frac{1}{k})-
\gamma_{bb}-\gamma_{hb}]}
{ \gamma_{hb}(1-\rho+\sum_{ij}\gamma_{ij}) } 
&=& \frac{1}{u}, \label{eq:fixed_hb}\\
\frac{(\frac{\rho_y}{k}-\gamma_{hh}-\gamma_{bh})[\rho_x (1-\frac{1}{k})-
\gamma_{bb}-\gamma_{bh}]}
{ \gamma_{bh}(1-\rho+\sum_{ij}\gamma_{ij}) } 
&=&\frac{1}{u},  \label{eq:fixed_bh}
\eea 
\end{subequations}
where $i,j = h,b$.
Equation~(\ref{eq:gamma}) can easily be solved to express 
$\gamma_{bb}$, $\gamma_{hb}$
and $\gamma_{bh}$ in 
terms of  $\gamma_{hh}$:
\be
\gamma_{bb}= (k-1)^2 \gamma_{hh},~\gamma_{bh}=\gamma_{hb}=
(k-1) \gamma_{hh},
\ee
and $\gamma_{hh}$ satisfies the quadratic equation
\be
\gamma_{hh}^2 -\gamma_{hh} \frac{\rho- \rho u-1}{k^2 (1-u)}
- \frac{u \rho_x\rho_y}{k^4 (1-u)}=0.
\label{eq:solve_bh}
\ee

From Eq.~(\ref{eq:partition}), the free energy is calculated using 
Eq.~(\ref{eq:free_defn}). We express the free energy in terms
of $\rho_x$, $\rho_y$ and $u$ as 
\begin{widetext}
\bea
f(\rho_x,\rho_y, u)& =&  
-\frac{k-1}{k}\sum_i \rho_i \ln\rho_i 
-\sum_i \left[1-\frac{(k-1)\rho_i}{k} \right]
\ln\left[1-\frac{(k-1)\rho_i}{k} \right]
+\sum_i (\rho_i - k^2 \gamma_{hh}) \ln (\rho_i - k^2 \gamma_{hh})
\nonumber \\ &&
+(1-\rho+k^2 \gamma_{hh}) \ln(1-\rho+k^2 \gamma_{hh})
-\frac{\rho}{k} \ln k
+k^2 \gamma_{hh} \ln \left(\frac{k^2 \gamma_{hh}}{u}\right),
\label{eq:free energy}
\eea
\end{widetext}
where $\gamma_{hh}$ is a function of $\rho_x$, $\rho_y$ and $u$ through
Eq.~(\ref{eq:solve_bh}). This expression for the free energy is not
convex everywhere. The true free energy $\bar{f}(\rho_x,\rho_y,u)$ is
obtained by the Maxwell construction such that
\be
\bar{f}(\rho_x,\rho_y,u)= \mathcal{CE} \left[ f(\rho_x,\rho_y,u)
\right],
\ee
where $\mathcal{CE}$ denotes the convex envelope.
Given total density $\rho$, the minimum of free energy 
determines $\rho_x$ and $\rho_y$.

\subsection{\label{subsec:transition_q_4}Two Phase Transitions}

\begin{figure}
\includegraphics[width=\columnwidth]{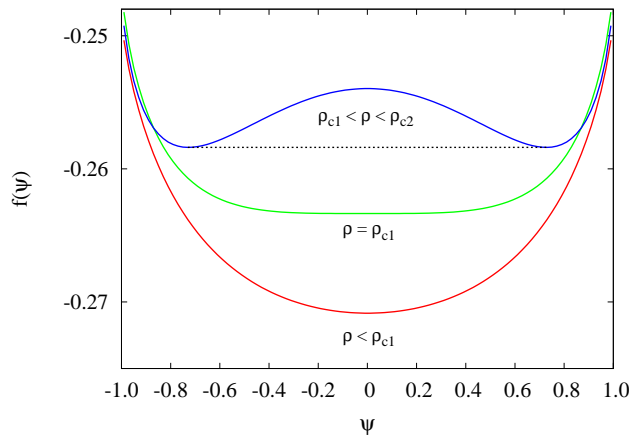}
\caption{Free energy $f(\psi)$ as a function of the order 
parameter $\psi$ for $\rho \approx \rho_{c1}$. The data are for  $k=6$,
$u=0.15$, and $q=4$. The curves have been shifted for clarity. 
The dotted line denotes the convex envelope.}
\label{fig:free_first}
\end{figure}

To study the phase transitions we define the nematic order parameter as 
\be
\psi=\frac{\rho_x-\rho_y}{\rho}.
\ee
The free energy can then be expressed as a power series in $\psi$,
\be
f(\rho_x,\rho_y,u) = A_0(\rho,u)+A_2 (\rho,u)\psi^2+
A_4(\rho,u)\psi^4+ \ldots,
\label{eq:landau}
\ee
where the coefficient $A_4(\rho,u) > 0$. $f(\rho_x,\rho_y,u)$ is 
unchanged when $\psi \leftrightarrow -\psi$. 
For small densities, the coefficient 
of the quadratic terms $A_2(\rho,u)$ is positive and the free energy
has a minimum at $\psi=0$ corresponding to the LDD phase. 
However for $k\geq4$, if $u$ is smaller 
than a critical value $u_c$, then $A_2(\rho,u)$ changes sign continuously at 
a critical density $\rho_{c1}$ and 
the free energy has two symmetric minima at $\psi\neq 0$, 
corresponding to the
nematic phase. This qualitative change in the behavior of the free
energy for densities
close to $\rho_{c1}$ is  shown in Fig.~\ref{fig:free_first}. 
As density is further increased, $A_2(\rho,u)$ changes sign continuously
from negative to positive at a second
critical density $\rho_{c2}$, such that the free energy has a minimum 
at $\psi=0$,
corresponding to the HDD phase. The  dependence of the free energy 
on $\psi$ for densities close to $\rho_{c2}$ is  similar to that shown
in Fig.~\ref{fig:free_first}.

\begin{figure} 
\includegraphics[width=\linewidth]{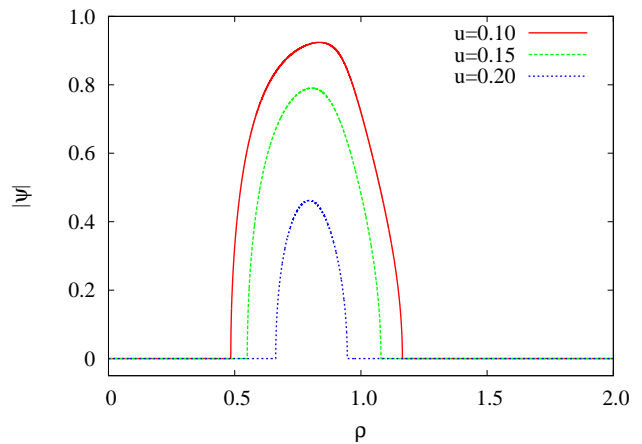} 
\caption{Order parameter $\psi$ as a function of density $\rho$. For low
and high densities,  $\psi = 0$,  while for intermediate densities, 
$\psi \neq 0$. The data are for $q=4$ and $k=6$.}
\label{fig:Q_rho} 
\end{figure}

\begin{figure}
\includegraphics[width=\linewidth]{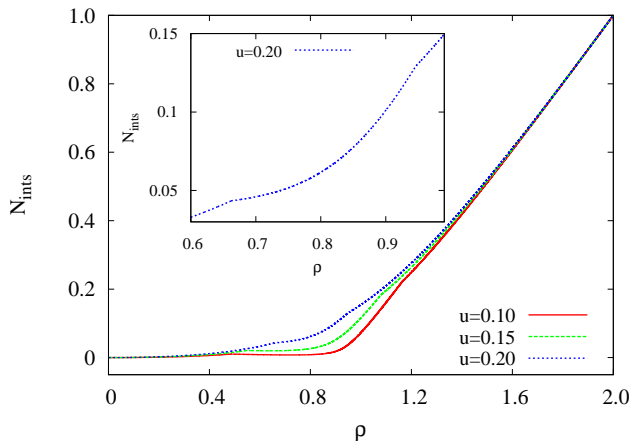} 
\caption{Average number of interactions per site, $N_{ints}$,
as a function of density $\rho$ for different values of $u$.  
Inset: The region between the two critical points is 
magnified. The data are for $q=4$ and $k=6$.}
\label{fig:fig_d_rho} 
\end{figure}

The variation of the order parameter $\psi$ with density $\rho$
is shown in Fig.~\ref{fig:Q_rho} for different values of $u$. 
$\psi$ increases continuously from
zero at $\rho_{c1}$ and decreases continuously to zero at $\rho_{c2}$.
The average number of intersections between the rods per site, 
though continuous, also shows non-analytic
behavior at $\rho_{c1}$ and $\rho_{c2}$ (see  
Fig.~\ref{fig:fig_d_rho}). 
The power series expansion of free energy in Eq.~(\ref{eq:landau}) has the
same form as that of a system with scalar order parameter that has
two broken symmetry phases. Thus, the two transitions will be 
in the mean field Ising universality class. 
The nematic phase does not exist for $k<4$.

\begin{figure}
\includegraphics[width=\linewidth]{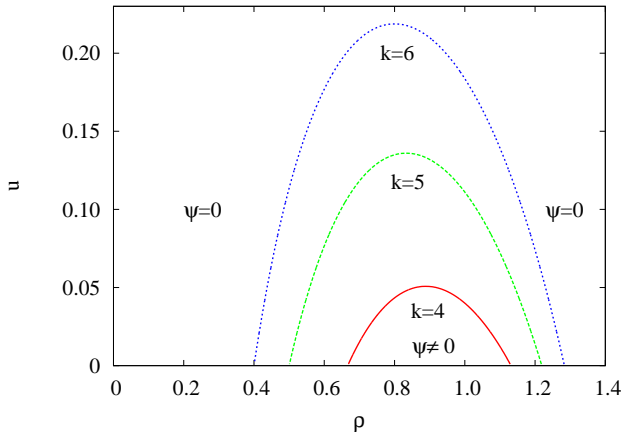} 
\caption{Phase diagram when $q=4$ for different values of $k$.}
\label{fig:fig_u_rho} 
\end{figure}

The phase diagram in the $\rho$--$u$ plane is determined by 
solving $A_2(\rho,u)=0$ for $\rho$ and is shown in
Fig.~\ref{fig:fig_u_rho} for different values of $k$.
The difference between the two critical densities 
decreases with increasing
$u$.  Beyond a maximum value $u_c(k)$,
there is no phase transition and 
the system remains disordered at all densities. The critical
densities $\rho_{c1}$ and $\rho_{c2}$  may be solved as an expansion
in $u$. For example, when $k=4$,  
\be
\rho_{c1} =\frac{2}{k-1}+ 2 u +12 u^{2} + O(u^3), ~~k=4.
\ee
and
\be
\rho_{c2}=1.13148-2.38675 u-12.2726 u^2+O(u^3),
~k=4.
\ee
It is of interest to determine $\rho_{c2} $ for large $k$. 
For the hard rod problem,
it was conjectured that $\rho_{c2}\approx 1- a/k^2$, 
when $k\rightarrow \infty$~\cite{ghosh2007}.
For our model, we find,
\bea
\rho_{c2} &=&\frac{-1+2 k-\sqrt{-3+4 k}}{-1+k}, ~~u \rightarrow 0,
\nonumber\\
&=&2-\frac{2}{\sqrt{k}}+\frac{1}{k} -\frac{5}{4
k^{3/2}}+\frac{1}{k^2}+
O(k^{-5/2}).
\eea
Thus the leading correction is $O(1/\sqrt{k})$, and not $O(1/k^2)$.

\begin{figure}
\includegraphics[width=\linewidth]{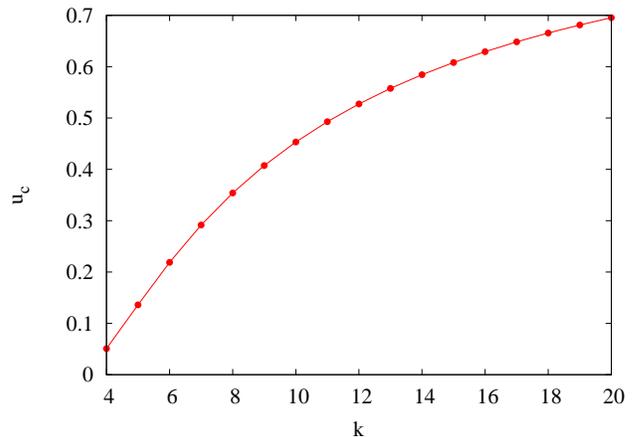} 
\caption{$u_c$, the maximum value of $u$ for which the transitions exists 
as a function of $k$. The data are for $q=4$.}
\label{fig:fig_uc_k} 
\end{figure}

$u_c(k)$, the largest value of $u$ for which the nematic phase
exists, is determined by solving the equations $A_2(\rho,u)=0$ and 
$d A_2(\rho,u)/d\rho =0$ simultaneously. $u_c(k)$ increases with $k$ (see
Fig.~\ref{fig:fig_uc_k}), and approaches $1$ from below as
$k\rightarrow \infty$.
At $u_c(k)$ two mean-field Ising critical lines meet. 

\section{\label{sec:q_6}$k$-mers  on RLTL with  $q=6$}

The calculation presented in Sec.~\ref{sec:q_4} may be extended to the case
when the coordination number $q \geq 6$.  We discuss the results when $q=6$. 
In this case, we associate a weight $u$ $(v)$ to a site occupied by
two (three) $k$-mers of different type.  
The calculation of the free energy now involves many more
combinatorial factors than for the case $q=4$, but is straightforward. 
The details of the calculation may be found in 
Supplementary material~\cite{supplement}.
Let $\rho_x$, $\rho_y$ and $\rho_z$
be the fraction of sites occupied by $x$-mers, $y$-mers and 
$z$-mers respectively. We define the order 
parameter to be 
$\psi=(\rho_x-\rho_y)/\rho$, where we set $\rho_y=\rho_z$. 
We find that for $u<u_c(k)$ and $v<u$, the system 
undergoes two transitions as for the case $q=4$, at
critical densities $\rho_{c1}$ and $\rho_{c2}$.

\begin{figure}
\includegraphics[width=\columnwidth]{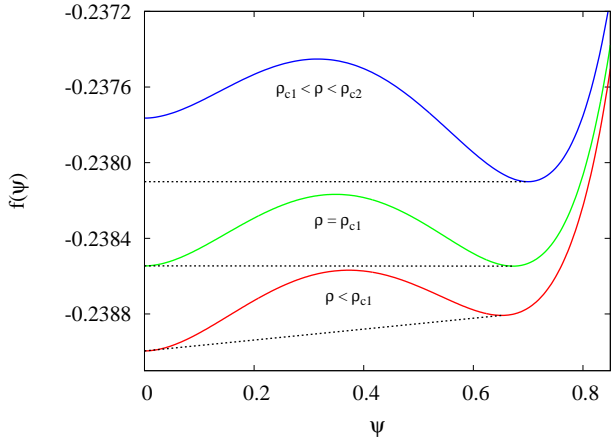}
\caption{Free energy $f(\psi)$ as a function of the order
parameter $\psi$ for $\rho \approx \rho_{c1}$ when $q=6$. The data are for 
$k=6$, $v=u^2$ and $u=0.15$. 
The dotted lines denote the convex envelopes.}
\label{fig:free_first_q6}
\end{figure}

The three dimensional $\rho$--$u$--$v$ phase diagram may be
visualized by studying the phase diagram along three different lines in the $u$--$v$ 
plane: $v=u^2$, $v=u^3$ and $v=u^4$. 
The free energy,  expressed as a power series in $\psi$, now has
the form
\bea
f(\rho_x,\rho_y,u,v)& =& A_0(\rho,u,v)+A_2 (\rho,u,v)\psi^2 \nonumber
\\
&+& A_3(\rho,u,v)
\psi^3+ A_4(\rho,u,v)\psi^4+ \ldots,
\label{eq:landauq6}
\eea
where $A_4(\rho,u,v) >0$ and $A_3(\rho,u,v)$ is in general non-zero.
At low densities, $A_2(\rho,u,v)$ is positive and the free energy has a 
global minimum at $\psi=0$.
With increasing density it develops a second local minimum at $\psi\neq 0$. 
At $\rho_{c1}$ the two minima become degenerate, and for $\rho_{c1}< 
\rho<\rho_{c2}$, the free energy has a minimum at
$\psi \neq 0$, corresponding 
to the nematic phase. A typical example
is shown in Fig.~\ref{fig:free_first_q6}.
The order parameter thus shows a discontinuity at $\rho_{c1}$ and the
transition is first order. 
In all the cases we have studied,  we find  that
the first transition from disordered to nematic phase is discontinuous.

\begin{figure}
\includegraphics[width=\linewidth]{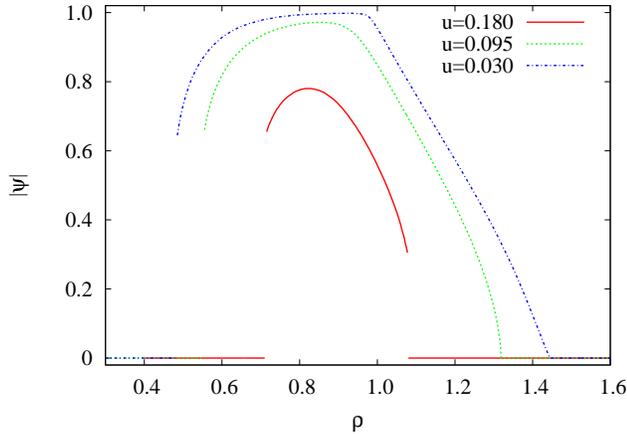} 
\caption{Order parameter $\psi$ as a function of 
density $\rho$ for different values of $u$ for $k=7$, $q=6$, and
$v=u^3$. The
second transition is first order for $u > u^*(k)$ and continuous for
$u \leq u^*(k)$. Here $u^*(7) \approx 0.09563$.
}
\label{fig:Q_rho_q6} 
\end{figure}

On the other hand, the nature of  the second transition from the nematic to HDD
phase depends on the value of $k$, $u$ and $v$. When  $v=u^2$, the second 
transition is first order for all $k$. However, when $v=u^3$, the
second transition could be first order or continuous. We find that for 
$k<7$, the second transition is always first order. For $k=7$, the
variation of the order parameter $\psi$ with density $\rho$ is shown
in Fig.~\ref{fig:Q_rho_q6}. Qualitatively similar behavior is seen for
$k>7$. The second transition is continuous for $u\leq u^*(k)$ and
first order for $u>u^*(k)$. The value of $u^*(k)$ increases with $k$.
When $v=u^4$, the phenomenology is
qualitatively similar to that for the case $v=u^3$.

\begin{figure}
\includegraphics[width=\linewidth]{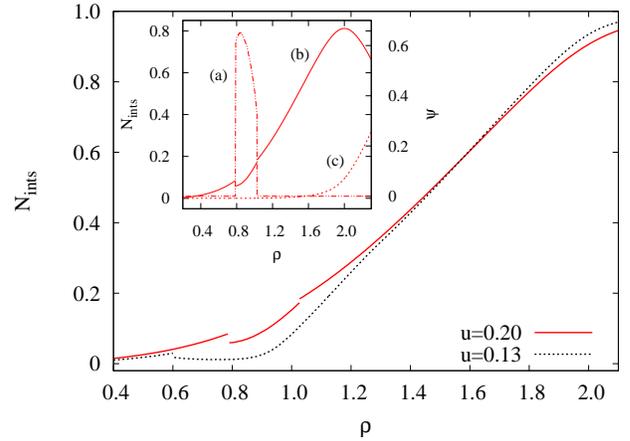} 
\caption{The number of interactions per site, $N_{ints}$, 
as a function of density $\rho$ for two different values of $u$. 
The data are for $q=6$, $k=7$, and $v=u^4$. 
Inset: The variation with density of (a) order parameter $\psi$, 
(b) fraction of sites occupied by two $k$-mers, and (c) fraction of
sites occupied by three $k$-mers. Here,
$u=0.20$.}
\label{fig:d_rho_q6} 
\end{figure}

The first order or continuous nature of the second transition is also
reflected in the average number of intersections. In
Fig.~\ref{fig:d_rho_q6}, we show the variation for 
the number of intersections per site with density for $k=7$ for two
values of $u$: one corresponding to a first order and the other to
continuous transition.  In addition to 
$\psi$, the average number of intersections between
rods per site also shows a discontinuity when the transition 
is first order.  This  discontinuity vanishes when the transition becomes 
continuous. 

\begin{figure}
\includegraphics[width=\linewidth]{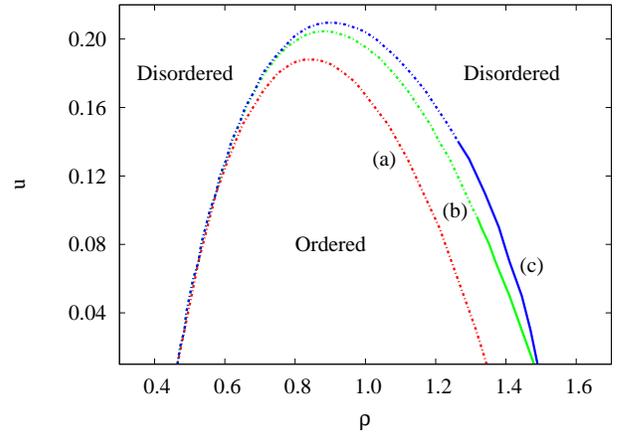} 
\caption{Phase diagram for $q=6$ and $k=7$ for (a)  $v=u^2$, 
(b) $v=u^3$, and (c) $v=u^4$. Dotted (solid) lines denotes first 
(second) order transitions.}
\label{fig:phase_diag_q6} 
\end{figure}

These observations are summarized in the $\rho$--$u$ phase diagram for
$k=7$ shown in Fig.~\ref{fig:phase_diag_q6}. For $v=u^3$ and $v=u^4$,
a second order line 
terminates at a tricritical point beyond which the transition becomes 
first order. 

\begin{figure}
\includegraphics[width=\linewidth]{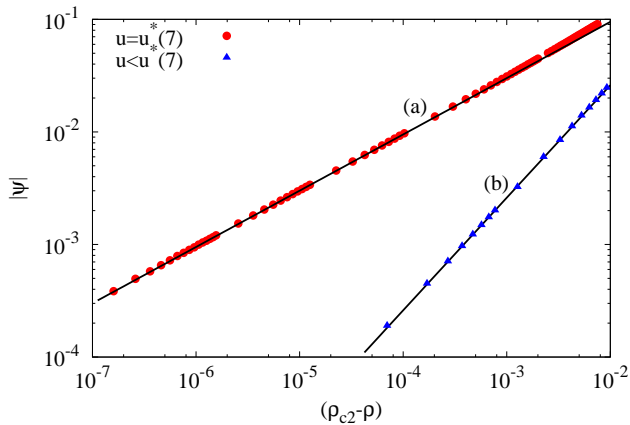} 
\caption{The order parameter $\psi$ as the density $\rho$ approaches
the critical density $\rho_{c2}$ for $u<u^*$ and at the tricritical
point $u=u^*$ when $k=7$,
$q=6$ and $v=u^3$. The solid lines are power laws (a)
$(\rho_{c2}-\rho)^{1/2}$ and (b) $(\rho_{c2}-\rho)$.
}
\label{fig:exponents} 
\end{figure}

The exponents describing the continuous transitions may
be found from the Landau-type free energy, Eq.~(\ref{eq:landauq6}).
At the first transition $A_2(\rho,u,v)>0$
and $A_3(\rho, u, v) <0$. At the spinodal point $A_2(\rho,u,v)$ changes
sign to negative. As density is further increased $A_2(\rho,u,v)$ changes
sign back to positive. When this occurs, 
$A_3(\rho,u,v)$ could be
positive or negative. If positive, then the transition will be
continuous. Now the critical exponents are determined from a Landau free 
energy functional
of the form $A_2 \psi^2 + A_3 \psi^3$, and hence the critical exponent 
$\beta = 1$, where  $\psi \sim (\rho_{c2}-\rho)^\beta$ as $\rho$
approaches $\rho_{c2}$ from below. 
At the tricritical point $A_3(\rho,u,v)=0$, and the
transition is in the mean field Ising universality class with
$\beta=1/2$ (see Fig.~\ref{fig:exponents}). 

\section{\label{sec:summary}Summary and Discussion}

In this paper, we studied the problem of monodispersed
long rigid rods on the RLTL, a Bethe-like lattice where rods of
different orientations are allowed to intersect 
with weight $u, v, \ldots$ depending on whether a site is occupied by
two, three, $\ldots$ $k$-mers.
We showed that the system undergoes two phase 
transitions with increasing density for  $k\geq k_{min}$  and
appropriate choice of interaction parameters.
For coordination number $q=4$, the two transitions
are continuous and in the mean field Ising universality class. 
For $q= 6$, 
while the first transition is first order, the nature of the 
second transition depends on the values  $k$, $u$ and $v$, giving rise
to a rich phase diagram. To the best of our knowledge,
it is the only solvable model on interacting rods that shows 
two phase transitions.

The limit $u\to 0$ is different 
from $u=0$ (the hard rod problem). When $u=0$,
the second transition in absent~\cite{rajesh2011}.
When $u, v>0$, the fully packed phase  is disordered by construction 
and if the first phase transition exists, so does a second phase transition.
The relaxation of the 
restriction that only rods of different orientations may
intersect at a lattice  site does not change the 
qualitative behavior
of the system as the high density phase remains disordered. 
There are still two transitions, both in the mean field
Ising universality class (when $q=4$). However, the solution becomes more
cumbersome. 

Similarly when $q=6$, the limit $v\rightarrow 0$ is different from $v=0$ when
$u >0$. When $v=0$, a lattice site may occupied by utmost two
$k$-mers of different type. In this case, the fully packed phase is
not necessarily disordered and for certain values of $k$ and $u$,
only one transition is present for increasing density.

For hard rods on the square lattice, Monte Carlo simulations were
unable to give a  clear answer to the question
whether the HDD and LDD phases are qualitatively similar or 
not~\cite{joyjit2013}. It was argued that the HDD phase has a large
crossover length scale $\xi^*$ $\sim 1500$, and 
for length scales larger than  $\xi^*$ 
it is possible that the HDD phase is not qualitatively different
from the LDD phase. This was
based on the evidence that vacancies in the HDD phase do not form a
bound state. In this paper, by expanding the phase diagram from a one-dimensional
$\rho$ phase diagram to a multi-dimensional $\rho$--interaction parameters phase diagram,
we showed that it is always possible to continuously transform the LDD phase into
the HDD phase without crossing any phase boundary. This means that the LDD
and HDD phases are qualitatively similar, at least for the model on RLTL.
It would thus be worthwhile to simulate
the hard rods problem on the 
square lattice for system sizes larger than $1500$ and verify the
same.

It would also be possible to study the problem with repulsive interactions on the
square lattice. The algorithm presented in Refs.~\cite{joyjit_dae,joyjit2013} is generalizable
to the case when intersections are allowed. Confirming whether the qualitative behavior is 
similar to that seen for RLTL would be interesting. Measuring the exponents for the second 
transition might be easier for such a model as the critical density would be away from the 
fully packed limit.

For the RLTL with coordination number $q=4$, we showed that for large $k$,
$\rho_{c2} \approx 2- a/\sqrt{k}+ O(k^{-1})$. This is at variance from the 
prediction from entropy based  arguments for the hard rod problem
that $\rho_{c2}$ approaches $1$ as $k^{-2}$~\cite{ghosh2007}.  It would be 
interesting to resolve this discrepancy.

The RLTL is suitable for studying problems that show orientational
order. Polydispersed systems can show multiple
phases~\cite{sollich2003_02,sollich2003_03}.
Its solution on the RLTL would make rigorous some of the qualitative
features of the problem. This is a promising area for
future study.

\section*{Acknowledgments}
We thank Deepak Dhar and J\"{u}rgen F. Stilck for very helpful discussions.

\appendix

\section{\label{sec:appendix1} Calculation of $C_m$ for $q=4$}

In this appendix, we derive the expression for $C_m$ in Eq.~(\ref{eq:cm}). $C_m$ is the
total number of ways of connecting the bonds from the $(m-1)^{th}$ layer to 
the $m^{th}$ layer 
consistent with the number of $x$-mers,  $y$-mers, and intersections at the $m^{th}$
layer.

In the $(m-1)^{th}$ layer, there are $X_m$ and $Y_m$ sites occupied 
by $x$-mers and $y$-mers that extend to the $m^{th}$ layer. 
These $X_m$ bonds of type 
$X$ can be connected to $X_m$ different sites in the $m^{th}$ layer in 
\[
\frac{N!}{(N-X_m)!}
\]
ways. Among the $Y_m$ bonds of type $Y$, $\Gamma^m_{bb}$ of them are 
connected to 
sites occupied by  an $x$-mer and the remaining $Y_m- \Gamma^m_{bb}$ bonds 
are connected to empty sites in the $m^{th}$ layer.
The number of ways of connecting is
\[
\frac{Y_m! X_m!}{\Gamma^m_{bb}! (Y-\Gamma^m_{bb})! (X_m-\Gamma^m_{bb})!}
\times \frac{(N-X_m)!}{(N-X_m-Y_m+\Gamma^m_{bb})!}.
\]

Now connect the remaining  $(N-X_m)$ free bonds of type $X$ and $(N-Y_m)$ 
free bonds of type $Y$  to sites in layer $m$ that are not occupied by 
$x$-mers and $y$-mers respectively. This can be done in 
\[
(N-X_m)! (N-Y_m)!
\]
ways. 

We have to now assign sites to $x_m$ and $y_m$ heads in layer $m$. Out
of $x_m$ ($y_m$) heads, $\Gamma_{hb}^m$ ($\Gamma_{bh}^m$)
of them will be on sites already
occupied by only a $y$-mer ($x$-mer). The number of ways of doing this
is
\[
\frac{(X_m-\Gamma^m_{bb})!}{\Gamma^m_{bh}!(X_m-\Gamma^m_{bb}-\Gamma^m_{bh})!}
\times
\frac{(Y_m-\Gamma^m_{bb})!}{\Gamma^m_{hb}!(Y_m-\Gamma^m_{bb}-\Gamma^m_{hb})!}.
\]

There are $(N-X_m-Y_m+\Gamma^m_{bb})$ sites in the $m^{th}$ layer which 
are unoccupied so far. They can be divided into four groups: 
$\Gamma^m_{hh}$ sites, each occupied by 
the heads of an $x$-mer and a $y$-mer,
$(x_m-\Gamma^m_{hh}-\Gamma^m_{hb})$ sites occupied by only a head of an
$x$-mer, $(y_m-\Gamma^m_{hh}-\Gamma^m_{bh})$ sites occupied by only a
head of a $y$-mer, and $(N-X_m-Y_m-x_m-y_m+ \sum_{ij} \Gamma^m_{ij})$ 
unoccupied sites. The number of 
ways of arranging them is: 
\[
\begin{split}
\frac{(N-X_m-Y_m+\Gamma^m_{bb})! }
{\Gamma^m_{hh}! (x_m-\Gamma^m_{hh}-\Gamma^m_{hb})! 
(y_m-\Gamma^m_{hh}-\Gamma^m_{bh})!} \\
\times \frac{1}{(N-X_m-Y_m-x_m-y_m+ \sum_{ij}\Gamma^m_{ij})!}.
\end{split}
\]
The product of all these factors gives 
$C_m$ as given in Eq.~(\ref{eq:cm}).


\begin{thebibliography}{25}%
\makeatletter
\providecommand \@ifxundefined [1]{%
 \@ifx{#1\undefined}
}%
\providecommand \@ifnum [1]{%
 \ifnum #1\expandafter \@firstoftwo
 \else \expandafter \@secondoftwo
 \fi
}%
\providecommand \@ifx [1]{%
 \ifx #1\expandafter \@firstoftwo
 \else \expandafter \@secondoftwo
 \fi
}%
\providecommand \natexlab [1]{#1}%
\providecommand \enquote  [1]{``#1''}%
\providecommand \bibnamefont  [1]{#1}%
\providecommand \bibfnamefont [1]{#1}%
\providecommand \citenamefont [1]{#1}%
\providecommand \href@noop [0]{\@secondoftwo}%
\providecommand \href [0]{\begingroup \@sanitize@url \@href}%
\providecommand \@href[1]{\@@startlink{#1}\@@href}%
\providecommand \@@href[1]{\endgroup#1\@@endlink}%
\providecommand \@sanitize@url [0]{\catcode `\\12\catcode `\$12\catcode
  `\&12\catcode `\#12\catcode `\^12\catcode `\_12\catcode `\%12\relax}%
\providecommand \@@startlink[1]{}%
\providecommand \@@endlink[0]{}%
\providecommand \url  [0]{\begingroup\@sanitize@url \@url }%
\providecommand \@url [1]{\endgroup\@href {#1}{\urlprefix }}%
\providecommand \urlprefix  [0]{URL }%
\providecommand \Eprint [0]{\href }%
\providecommand \doibase [0]{http://dx.doi.org/}%
\providecommand \selectlanguage [0]{\@gobble}%
\providecommand \bibinfo  [0]{\@secondoftwo}%
\providecommand \bibfield  [0]{\@secondoftwo}%
\providecommand \translation [1]{[#1]}%
\providecommand \BibitemOpen [0]{}%
\providecommand \bibitemStop [0]{}%
\providecommand \bibitemNoStop [0]{.\EOS\space}%
\providecommand \EOS [0]{\spacefactor3000\relax}%
\providecommand \BibitemShut  [1]{\csname bibitem#1\endcsname}%
\let\auto@bib@innerbib\@empty
\bibitem [{\citenamefont {Onsager}(1949)}]{onsager1949}%
  \BibitemOpen
  \bibfield  {author} {\bibinfo {author} {\bibfnamefont {L.}~\bibnamefont
  {Onsager}},\ }\href@noop {} {\bibfield  {journal} {\bibinfo  {journal} {Ann.
  N.Y. Acad. Sci.}\ }\textbf {\bibinfo {volume} {51}},\ \bibinfo {pages} {627}
  (\bibinfo {year} {1949})}\BibitemShut {NoStop}%
\bibitem [{\citenamefont {Flory}(1956)}]{flory1956a}%
  \BibitemOpen
  \bibfield  {author} {\bibinfo {author} {\bibfnamefont {P.~J.}\ \bibnamefont
  {Flory}},\ }\href@noop {} {\bibfield  {journal} {\bibinfo  {journal} {Proc.
  R. Soc.}\ }\textbf {\bibinfo {volume} {234}},\ \bibinfo {pages} {60}
  (\bibinfo {year} {1956})}\BibitemShut {NoStop}%
\bibitem [{\citenamefont {Zwanzig}(1963)}]{zwanzig1963}%
  \BibitemOpen
  \bibfield  {author} {\bibinfo {author} {\bibfnamefont {R.}~\bibnamefont
  {Zwanzig}},\ }\href@noop {} {\bibfield  {journal} {\bibinfo  {journal} {J.
  Chem. Phys.}\ }\textbf {\bibinfo {volume} {39}},\ \bibinfo {pages} {1714}
  (\bibinfo {year} {1963})}\BibitemShut {NoStop}%
\bibitem [{\citenamefont {Vroege}\ and\ \citenamefont
  {Lekkerkerker}(1992)}]{vroege1992}%
  \BibitemOpen
  \bibfield  {author} {\bibinfo {author} {\bibfnamefont {G.~J.}\ \bibnamefont
  {Vroege}}\ and\ \bibinfo {author} {\bibfnamefont {H.~N.~W.}\ \bibnamefont
  {Lekkerkerker}},\ }\href@noop {} {\bibfield  {journal} {\bibinfo  {journal}
  {Rep. Prog. Phys.}\ }\textbf {\bibinfo {volume} {55}},\ \bibinfo {pages}
  {1241} (\bibinfo {year} {1992})}\BibitemShut {NoStop}%
\bibitem [{\citenamefont {de~Gennes}\ and\ \citenamefont
  {Prost}(1993)}]{degennesBook}%
  \BibitemOpen
  \bibfield  {author} {\bibinfo {author} {\bibfnamefont {P.~G.}\ \bibnamefont
  {de~Gennes}}\ and\ \bibinfo {author} {\bibfnamefont {J.}~\bibnamefont
  {Prost}},\ }\href@noop {} {\emph {\bibinfo {title} {The Physics of Liquid
  Crystals}}}\ (\bibinfo  {publisher} {Oxford University Press},\ \bibinfo
  {address} {Oxford},\ \bibinfo {year} {1993})\BibitemShut {NoStop}%
\bibitem [{\citenamefont {Mermin}\ and\ \citenamefont
  {Wagner}(1966)}]{mermin1966}%
  \BibitemOpen
  \bibfield  {author} {\bibinfo {author} {\bibfnamefont {N.~D.}\ \bibnamefont
  {Mermin}}\ and\ \bibinfo {author} {\bibfnamefont {H.}~\bibnamefont
  {Wagner}},\ }\href@noop {} {\bibfield  {journal} {\bibinfo  {journal} {Phys.
  Rev. Lett.}\ }\textbf {\bibinfo {volume} {17}},\ \bibinfo {pages} {1133}
  (\bibinfo {year} {1966})}\BibitemShut {NoStop}%
\bibitem [{\citenamefont {Straley}(1971)}]{straley1971}%
  \BibitemOpen
  \bibfield  {author} {\bibinfo {author} {\bibfnamefont {J.~P.}\ \bibnamefont
  {Straley}},\ }\href@noop {} {\bibfield  {journal} {\bibinfo  {journal} {Phys.
  Rev. A}\ }\textbf {\bibinfo {volume} {4}},\ \bibinfo {pages} {675} (\bibinfo
  {year} {1971})}\BibitemShut {NoStop}%
\bibitem [{\citenamefont {Frenkel}\ and\ \citenamefont
  {Eppenga}(1985)}]{frenkel1985}%
  \BibitemOpen
  \bibfield  {author} {\bibinfo {author} {\bibfnamefont {D.}~\bibnamefont
  {Frenkel}}\ and\ \bibinfo {author} {\bibfnamefont {R.}~\bibnamefont
  {Eppenga}},\ }\href@noop {} {\bibfield  {journal} {\bibinfo  {journal} {Phys.
  Rev. A}\ }\textbf {\bibinfo {volume} {31}},\ \bibinfo {pages} {1776}
  (\bibinfo {year} {1985})}\BibitemShut {NoStop}%
\bibitem [{\citenamefont {Khandkar}\ and\ \citenamefont
  {Barma}(2005)}]{khandkar2005}%
  \BibitemOpen
  \bibfield  {author} {\bibinfo {author} {\bibfnamefont {M.~D.}\ \bibnamefont
  {Khandkar}}\ and\ \bibinfo {author} {\bibfnamefont {M.}~\bibnamefont
  {Barma}},\ }\href@noop {} {\bibfield  {journal} {\bibinfo  {journal} {Phys.
  Rev. E}\ }\textbf {\bibinfo {volume} {72}},\ \bibinfo {pages} {051717}
  (\bibinfo {year} {2005})}\BibitemShut {NoStop}%
\bibitem [{\citenamefont {Vink}(2009)}]{vink2009}%
  \BibitemOpen
  \bibfield  {author} {\bibinfo {author} {\bibfnamefont {R.~L.~C.}\
  \bibnamefont {Vink}},\ }\href@noop {} {\bibfield  {journal} {\bibinfo
  {journal} {Euro. Phys. J. B}\ }\textbf {\bibinfo {volume} {72}},\ \bibinfo
  {pages} {225} (\bibinfo {year} {2009})}\BibitemShut {NoStop}%
\bibitem [{\citenamefont {Ghosh}\ and\ \citenamefont {Dhar}(2007)}]{ghosh2007}%
  \BibitemOpen
  \bibfield  {author} {\bibinfo {author} {\bibfnamefont {A.}~\bibnamefont
  {Ghosh}}\ and\ \bibinfo {author} {\bibfnamefont {D.}~\bibnamefont {Dhar}},\
  }\href@noop {} {\bibfield  {journal} {\bibinfo  {journal} {Euro. Phys.
  Lett.}\ }\textbf {\bibinfo {volume} {78}},\ \bibinfo {pages} {20003}
  (\bibinfo {year} {2007})}\BibitemShut {NoStop}%
\bibitem [{\citenamefont {Disertori}\ and\ \citenamefont
  {Giuliani}(2012)}]{giuliani2012}%
  \BibitemOpen
  \bibfield  {author} {\bibinfo {author} {\bibfnamefont {M.}~\bibnamefont
  {Disertori}}\ and\ \bibinfo {author} {\bibfnamefont {A.}~\bibnamefont
  {Giuliani}},\ }\href@noop {} {\bibfield  {journal} {\bibinfo  {journal}
  {arXiv:1112.5564, to appear on Comm. Math. Phys.}\ } (\bibinfo {year}
  {2013})}\BibitemShut {NoStop}%
\bibitem [{\citenamefont {Kundu}\ \emph {et~al.}(2013)\citenamefont {Kundu},
  \citenamefont {Rajesh}, \citenamefont {Dhar},\ and\ \citenamefont
  {Stilck}}]{joyjit2013}%
  \BibitemOpen
  \bibfield  {author} {\bibinfo {author} {\bibfnamefont {J.}~\bibnamefont
  {Kundu}}, \bibinfo {author} {\bibfnamefont {R.}~\bibnamefont {Rajesh}},
  \bibinfo {author} {\bibfnamefont {D.}~\bibnamefont {Dhar}}, \ and\ \bibinfo
  {author} {\bibfnamefont {J.~F.}\ \bibnamefont {Stilck}},\ }\href@noop {}
  {\bibfield  {journal} {\bibinfo  {journal} {Phys. Rev. E}\ }\textbf {\bibinfo
  {volume} {87}},\ \bibinfo {pages} {032103} (\bibinfo {year}
  {2013})}\BibitemShut {NoStop}%
\bibitem [{\citenamefont {Heilmann}\ and\ \citenamefont
  {Lieb}(1972)}]{lieb1972}%
  \BibitemOpen
  \bibfield  {author} {\bibinfo {author} {\bibfnamefont {O.~J.}\ \bibnamefont
  {Heilmann}}\ and\ \bibinfo {author} {\bibfnamefont {E.}~\bibnamefont
  {Lieb}},\ }\href@noop {} {\bibfield  {journal} {\bibinfo  {journal} {Commun.
  Math. Phys.}\ }\textbf {\bibinfo {volume} {25}},\ \bibinfo {pages} {190}
  (\bibinfo {year} {1972})}\BibitemShut {NoStop}%
\bibitem [{\citenamefont {Matoz-Fernandez}\ \emph
  {et~al.}(2008{\natexlab{a}})\citenamefont {Matoz-Fernandez}, \citenamefont
  {Linares},\ and\ \citenamefont {Ramirez-Pastor}}]{fernandez2008c}%
  \BibitemOpen
  \bibfield  {author} {\bibinfo {author} {\bibfnamefont {D.~A.}\ \bibnamefont
  {Matoz-Fernandez}}, \bibinfo {author} {\bibfnamefont {D.~H.}\ \bibnamefont
  {Linares}}, \ and\ \bibinfo {author} {\bibfnamefont {A.~J.}\ \bibnamefont
  {Ramirez-Pastor}},\ }\href@noop {} {\bibfield  {journal} {\bibinfo  {journal}
  {J. Chem. Phys.}\ }\textbf {\bibinfo {volume} {128}},\ \bibinfo {pages}
  {214902} (\bibinfo {year} {2008}{\natexlab{a}})}\BibitemShut {NoStop}%
\bibitem [{\citenamefont {Matoz-Fernandez}\ \emph
  {et~al.}(2008{\natexlab{b}})\citenamefont {Matoz-Fernandez}, \citenamefont
  {Linares},\ and\ \citenamefont {Ramirez-Pastor}}]{fernandez2008a}%
  \BibitemOpen
  \bibfield  {author} {\bibinfo {author} {\bibfnamefont {D.~A.}\ \bibnamefont
  {Matoz-Fernandez}}, \bibinfo {author} {\bibfnamefont {D.~H.}\ \bibnamefont
  {Linares}}, \ and\ \bibinfo {author} {\bibfnamefont {A.~J.}\ \bibnamefont
  {Ramirez-Pastor}},\ }\href@noop {} {\bibfield  {journal} {\bibinfo  {journal}
  {Euro. Phys. Lett}\ }\textbf {\bibinfo {volume} {82}},\ \bibinfo {pages}
  {50007} (\bibinfo {year} {2008}{\natexlab{b}})}\BibitemShut {NoStop}%
\bibitem [{\citenamefont {Matoz-Fernandez}\ \emph
  {et~al.}(2008{\natexlab{c}})\citenamefont {Matoz-Fernandez}, \citenamefont
  {Linares},\ and\ \citenamefont {Ramirez-Pastor}}]{fernandez2008b}%
  \BibitemOpen
  \bibfield  {author} {\bibinfo {author} {\bibfnamefont {D.~A.}\ \bibnamefont
  {Matoz-Fernandez}}, \bibinfo {author} {\bibfnamefont {D.~H.}\ \bibnamefont
  {Linares}}, \ and\ \bibinfo {author} {\bibfnamefont {A.~J.}\ \bibnamefont
  {Ramirez-Pastor}},\ }\href@noop {} {\bibfield  {journal} {\bibinfo  {journal}
  {Physica A}\ }\textbf {\bibinfo {volume} {387}},\ \bibinfo {pages} {6513}
  (\bibinfo {year} {2008}{\natexlab{c}})}\BibitemShut {NoStop}%
\bibitem [{\citenamefont {Linares}\ \emph {et~al.}(2008)\citenamefont
  {Linares}, \citenamefont {Rom\'{a}},\ and\ \citenamefont
  {Ramirez-Pastor}}]{linares2008}%
  \BibitemOpen
  \bibfield  {author} {\bibinfo {author} {\bibfnamefont {D.~H.}\ \bibnamefont
  {Linares}}, \bibinfo {author} {\bibfnamefont {F.}~\bibnamefont {Rom\'{a}}}, \
  and\ \bibinfo {author} {\bibfnamefont {A.~J.}\ \bibnamefont
  {Ramirez-Pastor}},\ }\href@noop {} {\bibfield  {journal} {\bibinfo  {journal}
  {J. Stat. Mech.}\ ,\ \bibinfo {pages} {P03013}} (\bibinfo {year}
  {2008})}\BibitemShut {NoStop}%
\bibitem [{\citenamefont {Fischer}\ and\ \citenamefont
  {Vink}(2009)}]{fischer2009}%
  \BibitemOpen
  \bibfield  {author} {\bibinfo {author} {\bibfnamefont {T.}~\bibnamefont
  {Fischer}}\ and\ \bibinfo {author} {\bibfnamefont {R.~L. C.~.}\ \bibnamefont
  {Vink}},\ }\href@noop {} {\bibfield  {journal} {\bibinfo  {journal} {Euro.
  Phys. Lett.}\ }\textbf {\bibinfo {volume} {85}},\ \bibinfo {pages} {56003}
  (\bibinfo {year} {2009})}\BibitemShut {NoStop}%
\bibitem [{\citenamefont {Kundu}\ \emph {et~al.}(2012)\citenamefont {Kundu},
  \citenamefont {Rajesh}, \citenamefont {Dhar},\ and\ \citenamefont
  {Stilck}}]{joyjit_dae}%
  \BibitemOpen
  \bibfield  {author} {\bibinfo {author} {\bibfnamefont {J.}~\bibnamefont
  {Kundu}}, \bibinfo {author} {\bibfnamefont {R.}~\bibnamefont {Rajesh}},
  \bibinfo {author} {\bibfnamefont {D.}~\bibnamefont {Dhar}}, \ and\ \bibinfo
  {author} {\bibfnamefont {J.~F.}\ \bibnamefont {Stilck}},\ }\href@noop {}
  {\bibfield  {journal} {\bibinfo  {journal} {AIP Conf. Proc.}\ }\textbf
  {\bibinfo {volume} {1447}},\ \bibinfo {pages} {113} (\bibinfo {year}
  {2012})}\BibitemShut {NoStop}%
\bibitem [{\citenamefont {Dhar}\ \emph {et~al.}(2011)\citenamefont {Dhar},
  \citenamefont {Rajesh},\ and\ \citenamefont {Stilck}}]{rajesh2011}%
  \BibitemOpen
  \bibfield  {author} {\bibinfo {author} {\bibfnamefont {D.}~\bibnamefont
  {Dhar}}, \bibinfo {author} {\bibfnamefont {R.}~\bibnamefont {Rajesh}}, \ and\
  \bibinfo {author} {\bibfnamefont {J.~F.}\ \bibnamefont {Stilck}},\
  }\href@noop {} {\bibfield  {journal} {\bibinfo  {journal} {Phys. Rev. E}\
  }\textbf {\bibinfo {volume} {84}},\ \bibinfo {pages} {011140} (\bibinfo
  {year} {2011})}\BibitemShut {NoStop}%
\bibitem{supplement}
see Supplementary Material
\bibitem [{\citenamefont {Speranza}\ and\ \citenamefont
  {Sollich}(2003{\natexlab{b}})}]{sollich2003_02}%
  \BibitemOpen
  \bibfield  {author} {\bibinfo {author} {\bibfnamefont {A.}~\bibnamefont
  {Speranza}}\ and\ \bibinfo {author} {\bibfnamefont {P.}~\bibnamefont
  {Sollich}},\ }\href@noop {} {\bibfield  {journal} {\bibinfo  {journal} {Phys.
  Rev. E}\ }\textbf {\bibinfo {volume} {67}},\ \bibinfo {pages} {061702}
  (\bibinfo {year} {2003}{\natexlab{b}})}\BibitemShut {NoStop}%
\bibitem [{\citenamefont {Fasolo}\ and\ \citenamefont
  {Sollic}(2003)}]{sollich2003_03}%
  \BibitemOpen
  \bibfield  {author} {\bibinfo {author} {\bibfnamefont {M.}~\bibnamefont
  {Fasolo}}\ and\ \bibinfo {author} {\bibfnamefont {P.}~\bibnamefont
  {Sollich}},\ }\href@noop {} {\bibfield  {journal} {\bibinfo  {journal} {Phys.
  Rev. Lett.}\ }\textbf {\bibinfo {volume} {91}},\ \bibinfo {pages} {068301}
  (\bibinfo {year} {2003})}\BibitemShut {NoStop}%
\end{thebibliography}

%

\end{document}


\title{{\it Supplementary Information for} 
Re-entrant Disordered Phase in a  System of Repulsive Rods
on a Bethe-like Lattice}
\author{Joyjit Kundu}
\author{R. Rajesh}

\maketitle

\section{Calculation of free energy for coordination number $q=6$}

The combinatorial factor $C_m$
for $q=6$ may be obtained by following the same steps as in the case
$q=4$.  For $q=6$ a site can be occupied by utmost three
$k$-mers of different types. 
Let $\Gamma^{pq}_{nl}$ be the total number of intersections at the
$m^{th}$ layer
between site $n$ of a $k$-mer of type $p$ and site $l$ of a $k$-mer
of type $q$. $\Gamma^{pqr}_{nlm}$ denotes the total number of sites 
at the $m^{th}$ layer shared by site $n$ of a $k$-mer of type $p$, 
site $l$ of a $k$-mer 
of type $q$ and site $m$ of a $k$-mer of type $r$. Here $p$, $q$ and 
$r$ can be $x$, $y$ or $z$.  $n$, $l$ and $m$ can be $h$ or $b$
depending on whether the site is the head or part of the body of the
$k$-mer.
We omit the layer index $m$ from $\Gamma$ for notational
simplicity.
In addition to $X_m$ and $Y_m$ defined in Eq.~(3) of the paper, we
define
\[
 Z_m= \displaystyle\sum_{j=1}^{k-1}z_{m-j},
\]
as the number of sites at the $m^{th}$ layer occupied by $z$-mers, 
extended from $(m-1)^{th}$ layer. Let $e^{\mu_3}$ be the weight 
associated with a $z$-mer. 
$X_m$ number of $x$-mers from $(m-1)^{th}$ layer are connected to the 
$X_m$ sites of $m^{th}$ layer through $X$ type bonds in
\[
\frac{N!}{(N-X_m)!}
\]
ways. After connecting the $x$-mers, we connect $y$-mers between $m^{th}$ and 
$(m-1)^{th}$ layer. Among the $Y_m$ number of $y$-mers, $\bar{\Gamma}^{xy}_{bb}$ of them are 
connected to the sites at the $m^{th}$ layer, which are already occupied by $x$-mers. Later on 
while connecting $z$-mers, some of the 
sites among $\bar{\Gamma}^{xy}_{bb}$ sites might be occupied by $z$ mers also. 
Thus we have $\bar{\Gamma}^{xy}_{bb}=\Gamma^{xy}_{bb}+\Gamma^{xyz}_{bbb}+
\Gamma^{zxy}_{hbb}$. Remaining $(Y_m-\bar{\Gamma}^{xy}_{bb})$ sites are 
connected to the empty sites of the $m^{th}$ layer. The number 
of ways of doing this is
\[
 \frac{Y_m! X_m!}{\bar{\Gamma}^{xy}_{bb}! (Y_m-\bar{\Gamma}^{xy}_{bb})! (X_m-\bar{\Gamma}^{xy}_{bb})!}
\times \frac{(N-X_m)!}{(N-X_m-Y_m+\bar{\Gamma}^{xy}_{bb})!}.
\]
Now we connect the $z$ mers. 
$\bar{\Gamma}^{xz}_{bb}$ and $\bar{\Gamma}^{yz}_{bb}$ number of 
$z$-mers from $(m-1)^{th}$ layer are connected with 
the sites at the $m^{th}$ layer which are occupied by $x$-mers and $y$-mers respectively. Here 
$\bar{\Gamma}^{xz}_{bb}=\Gamma^{xz}_{bb}+\Gamma^{yxz}_{hbb}$ and similarly, 
$\bar{\Gamma}^{yz}_{bb}=\Gamma^{yz}_{bb}+\Gamma^{xyz}_{hbb}$. $\Gamma^{xyz}_{bbb}$ 
number of $z$-mers are connected to the sites which are already simultaneously shared by 
$x$-mers and $y$-mers at the $m^{th}$ layer. Rest of the $(Z_m-\bar{\Gamma}^{xz}_{bb}-
\bar{\Gamma}^{yz}_{bb}-\Gamma^{xyz}_{bbb})$ number of $z$-mers are connected 
to the remaining empty sites of the $m^{th}$ layer. The number of ways of connecting them is 
\[
\begin{split} 
\frac{Z_m!}{\bar{\Gamma}^{xz}_{bb}!(Z_m-\bar{\Gamma}^{xz}_{bb})!}
\times \frac{(X_m-\bar{\Gamma}^{xy}_{bb})!}
{(X_m-\bar{\Gamma}^{xy}_{bb}-\bar{\Gamma}^{xz}_{bb})!} \\
\frac{(Z_m-\bar{\Gamma}^{xz}_{bb})!}{\bar{\Gamma}^{yz}_{bb}!(Z_m-\bar{\Gamma}^{xz}_{bb}-\bar{\Gamma}^{yz}_{bb})!}
\times \frac{(Y_m-\bar{\Gamma}^{xy}_{bb})!}{(Y_m-\bar{\Gamma}^{xy}_{bb}-\bar{\Gamma}^{yz}_{bb})!} \\
\times \frac{(Z_m-\bar{\Gamma}^{xz}_{bb}-\bar{\Gamma}^{yz}_{bb})!}{\Gamma^{xyz}_{bbb}!(Z_m-\bar{\Gamma}^{xz}_{bb}-
\bar{\Gamma}^{yz}_{bb}-\Gamma^{xyz}_{bbb})!} \times \frac{\bar{\Gamma}^{xy}_{bb}!}{(\bar{\Gamma}^{xy}_{bb}-\Gamma^{xyz}_{bbb})!}\\
\times \frac{(N-X_m-Y_m+\bar{\Gamma}^{xy}_{bb})!}{(N-X_m-Y_m-Z_m+\bar{\Gamma}^{xy}_{bb}+\bar{\Gamma}^{yz}_{bb}+\bar{\Gamma}^{xz}_{bb}+\Gamma^{xyz}_{bbb})!}.
\end{split}
\]
We can connect $(N-X_m)$, $(N-Y_m)$ and $(N-Z_m)$ free bonds of type $X$, $Y$ 
and $Z$ respectively to the empty sites at the $m^{th}$ layer in 
\[
 (N-X_m)!(N-Y_m)!(N-Z_m)!
\]
ways. 
Now we consider the $k$-mers, starting from the sites at the $m^{th}$ layer 
which already occupied by the $k$-mers extended from the previous layer. Number 
of ways of choosing the heads of the $k$-mers at the $m^{th}$ layer from 
the sites which are already occupied by two different $k$-mers is given by, 
\[
\frac{\bar{\Gamma}^{yz}_{bb}!}{\Gamma^{xyz}_{hbb}!(\bar{\Gamma}^{yz}_{bb}-\Gamma^{xyz}_{hbb})!} 
\times \frac{\bar{\Gamma}^{xz}_{bb}!}{\Gamma^{yxz}_{hbb}!(\bar{\Gamma}^{xz}_{bb}-\Gamma^{yxz}_{hbb})!} 
\times \frac{(\bar{\Gamma}^{xy}_{bb}-\Gamma^{xyz}_{bbb})!}{\Gamma^{zxy}_{hbb}!(\bar{\Gamma}^{xy}_{bb}-
\Gamma^{xyz}_{bbb}-\Gamma^{zxy}_{hbb})!}. 
\]
Similarly two $k$-mers may start from the same site. We can choose such pairs of 
heads at the $m^{th}$ layer from the sites having 
single $k$-mers passing through them in 
\[
\begin{split} 
 \frac{(Z_m-\bar{\Gamma}^{xz}_{bb}-\bar{\Gamma}^{yz}_{bb}-\Gamma^{xyz}_{bbb})!)}{\Gamma^{xyz}_{hhb}!
(Z_m-\bar{\Gamma}^{xz}_{bb}-\bar{\Gamma}^{yz}_{bb}-\Gamma^{xyz}_{bbb}-\Gamma^{xyz}_{hhb})!} 
\times \frac{(Y_m-\bar{\Gamma}^{xy}_{bb}-\bar{\Gamma}^{yz}_{bb})!}{\Gamma^{xzy}_{hhb}!
(Y_m-\bar{\Gamma}^{xy}_{bb}-\bar{\Gamma}^{yz}_{bb}-\Gamma^{xzy}_{hhb})!} \\
\times \frac{(X_m-\bar{\Gamma}^{xy}_{bb}-\bar{\Gamma}^{xz}_{bb})!}{\Gamma^{yzx}_{hhb}!
(X_m-\bar{\Gamma}^{xy}_{bb}-\bar{\Gamma}^{xz}_{bb}-\Gamma^{yzx}_{hhb})!}
\end{split}
\]
ways. 
The total number of $x$-mers, $y$-mers and $z$-mers starting from the $m^{th}$ 
layer are $x_m$, $y_m$ and $z_m$ respectively. 
Number of ways of starting $k$-mers from the sites, already 
occupied by single $k$-mers at the $m^{th}$ layer is, 
\[
\begin{split}
 \frac{(Y_m-\bar{\Gamma}^{xy}_{bb}-\bar{\Gamma}^{yz}_{bb}-\Gamma^{xzy}_{hhb})!}{\Gamma^{xy}_{hb}!
(Y_m-\bar{\Gamma}^{xy}_{bb}-\bar{\Gamma}^{yz}_{bb}-\Gamma^{xzy}_{hhb}-\Gamma^{xy}_{hb})!} \times 
\frac{(Z_m-\bar{\Gamma}^{xz}_{bb}-\bar{\Gamma}^{yz}_{bb}-\Gamma^{xyz}_{hhb}-\Gamma^{xyz}_{bbb})!}
{\Gamma^{xz}_{hb}!(Z_m-\bar{\Gamma}^{xz}_{bb}-\bar{\Gamma}^{yz}_{bb}-
\Gamma^{xyz}_{hhb}-\Gamma^{xyz}_{bbb}-\Gamma^{xz}_{hb})!}\\ \times 
\frac{(X_m-\bar{\Gamma}^{xy}_{bb}-\bar{\Gamma}^{xz}_{bb}-\Gamma^{yzx}_{hhb})!}{\Gamma^{yx}_{hb}!
(X_m-\bar{\Gamma}^{xy}_{bb}-\bar{\Gamma}^{xz}_{bb}-\Gamma^{yzx}_{hhb}-\Gamma^{yx}_{hb})!} \times 
\frac{(Z_m-\bar{\Gamma}^{xz}_{bb}-\bar{\Gamma}^{yz}_{bb}-\Gamma^{xyz}_{hhb}-\Gamma^{xyz}_{bbb}-\Gamma^{xz}_{hb})!}
{\Gamma^{yz}_{hb}!(Z_m-\bar{\Gamma}^{xz}_{bb}-\bar{\Gamma}^{yz}_{bb}-
\Gamma^{xyz}_{hhb}-\Gamma^{xyz}_{bbb}-\Gamma^{xz}_{hb}-\Gamma^{yz}_{hb})!}\\ \times 
\frac{(X_m-\bar{\Gamma}^{xy}_{bb}-\bar{\Gamma}^{xz}_{bb}-\Gamma^{yzx}_{hhb}-\Gamma^{yx}_{hb})!}
{\Gamma^{zx}_{hb}!(X_m-\bar{\Gamma}^{xy}_{bb}-\bar{\Gamma}^{xz}_{bb}-
\Gamma^{yzx}_{hhb}-\Gamma^{yx}_{hb}-\Gamma^{zx}_{hb})!} \times 
\frac{(Y_m-\bar{\Gamma}^{xy}_{bb}-\bar{\Gamma}^{yz}_{bb}-\Gamma^{xzy}_{hhb}-\Gamma^{xy}_{hb})!}
{\Gamma^{zy}_{hb}!(Y_m-\bar{\Gamma}^{xy}_{bb}-\bar{\Gamma}^{yz}_{bb}-\Gamma^{xzy}_{hhb}-
\Gamma^{xy}_{hb}-\Gamma^{zy}_{hb})!}.
\end{split}
\]

There are $(N-X_m-Y_m-Z_m+\bar{\Gamma}^{xy}_{bb}+
\bar{\Gamma}^{xz}_{bb}+\bar{\Gamma}^{yz}_{bb}+\Gamma^{xyz}_{bbb})$ 
unoccupied sites so far at the $m^{th}$ 
layer. They can be divided into four groups: sites shared simultaneously 
by two heads of different types of $k$-mers, sites occupied by three heads of 
the $k$-mers of different types, sites from which single $k$-mers start, and fully empty 
sites. Number of ways to arrange them is, 
\[
\begin{split}
\frac{(N-X_m-Y_m-Z_m+\bar{\Gamma}^{xy}_{bb}+
\bar{\Gamma}^{xz}_{bb}+\bar{\Gamma}^{yz}_{bb}+\Gamma^{xyz}_{bbb})!}
{\Gamma^{xy}_{hh}!\Gamma^{xz}_{hh}!\Gamma^{yz}_{hh}!\Gamma^{xyz}_{hhh}!N_{em}!
(x_m-\Gamma^{xy}_{hh}-\Gamma^{xz}_{hh}-\Gamma^{xy}_{hb}-\Gamma^{xz}_{hb}-
\Gamma^{xyz}_{hhb}-\Gamma^{xzy}_{hhb}-\Gamma^{xyz}_{hbb}-\Gamma^{xyz}_{hhh})!} \\ \times 
\frac{1}{(y_m-\Gamma^{xy}_{hh}-\Gamma^{yz}_{hh}-\Gamma^{yx}_{hb}-\Gamma^{yz}_{hb}-
\Gamma^{xyz}_{hhb}-\Gamma^{yzx}_{hhb}-\Gamma^{yxz}_{hbb}-\Gamma^{xyz}_{hhh})!}\\ \times 
\frac{1}{(z_m-\Gamma^{yz}_{hh}-\Gamma^{xz}_{hh}-\Gamma^{zx}_{hb}-\Gamma^{zy}_{hb}-
\Gamma^{xzy}_{hhb}-\Gamma^{yzx}_{hhb}-\Gamma^{zxy}_{hbb}-\Gamma^{xyz}_{hhh})!},
\end{split}
\]
where,
\begin{align} 
N_{em}&=N-X_m-x_m-Y_m-y_m-Z_m-z_m+
\frac{1}{2}\displaystyle \sum_{i,j,i\neq j} \left[\Gamma^{ij}_{hh}+
\bar{\Gamma}^{ij}_{bb}+2 \Gamma^{ij}_{hb}\right]  \nonumber \\ 
&+ \Gamma^{xyz}_{bbb}+2\left[\Gamma^{xyz}_{hhb}
+\Gamma^{xzy}_{hhb}+\Gamma^{yzx}_{hhb}+\Gamma^{xyz}_{hhh}\right]+\Gamma^{xyz}_{hbb}+\Gamma^{yxz}_{hbb}
+\Gamma^{zxy}_{hbb}. \nonumber
\end{align}
$[\bar{\Gamma}^{ij}_{bb}]$ and $[\Gamma^{ij}_{hh}]$ are symmetric in $i$ and $j$, 
but $[\Gamma^{ij}_{hb}]$ is not.
Multiplying all these factors and writing $\bar{\Gamma}$ in 
terms of $\Gamma$ we obtain $C_m$ 
for $q=6$. Define homogeneous, $N$-independent 
variables: $\gamma^{ij}_{nl}=\Gamma^{ij}_{nl}/N$, $\gamma^{ijk}_{nlm}=
\Gamma^{ijk}_{nlm}/N$, 
$\rho_x=x_m k/N$, $\rho_y=y_m k/N $ and $\rho_z=z_m k/N$.

Define, 
\begin{align}
F&=\rho_x-\frac{\rho_x}{k}-\gamma^{xy}_{bb}-\gamma^{xz}_{bb}-\gamma^{yx}_{hb}-\gamma^{zx}_{hb}
-\gamma^{zxy}_{hbb}-\gamma^{yxz}_{hbb}-\gamma^{yzx}_{hhb}-\gamma^{xyz}_{bbb}, \nonumber \\
G&=\rho_y-\frac{\rho_y}{k}-\gamma^{xy}_{bb}-\gamma^{yz}_{bb}-\gamma^{xy}_{hb}-\gamma^{zy}_{hb}
-\gamma^{zxy}_{hbb}-\gamma^{xyz}_{hbb}-\gamma^{xzy}_{hhb}-\gamma^{xyz}_{bbb}, \nonumber \\
W&=\rho_z-\frac{\rho_z}{k}-\gamma^{xz}_{bb}-\gamma^{yz}_{bb}-\gamma^{xz}_{hb}-\gamma^{yz}_{hb}
-\gamma^{yxz}_{hbb}-\gamma^{xyz}_{hbb}-\gamma^{xyz}_{hhb}-\gamma^{xyz}_{bbb}, \nonumber \\
f&=\frac{\rho_x}{k}-\gamma^{xy}_{hh}-\gamma^{xz}_{hh}-\gamma^{xy}_{hb}-\gamma^{xz}_{hb}
-\gamma^{xyz}_{hbb}-\gamma^{xyz}_{hhb}-\gamma^{xzy}_{hhb}-\gamma^{xyz}_{hhh}, \nonumber \\
g&=\frac{\rho_y}{k}-\gamma^{xy}_{hh}-\gamma^{yz}_{hh}-\gamma^{yx}_{hb}-\gamma^{yz}_{hb}
-\gamma^{yxz}_{hbb}-\gamma^{xyz}_{hhb}-\gamma^{yzx}_{hhb}-\gamma^{xyz}_{hhh}, \nonumber \\
w&=\frac{\rho_z}{k}-\gamma^{xz}_{hh}-\gamma^{yz}_{hh}-\gamma^{zy}_{hb}-\gamma^{zx}_{hb}
-\gamma^{zxy}_{hbb}-\gamma^{xzy}_{hhb}-\gamma^{yzx}_{hhb}-\gamma^{xyz}_{hhh}, \nonumber \\
D&=1-\rho+\frac{1}{2} \displaystyle\sum_{i\neq j}
\left[\gamma^{ij}_{bb}+\gamma^{ij}_{hh}+2 \gamma^{ij}_{hb}\right]+2\left[\gamma^{xyz}_{bbb}+\gamma^{xyz}_{hhh}
+\gamma^{xyz}_{hbb}+\gamma^{yxz}_{hbb}\right. \nonumber \\ &+ \left.\gamma^{zxy}_{hbb}+\gamma^{xyz}_{hhb}+\gamma^{xzy}_{hhb}
+\gamma^{yzx}_{hhb}\right]. \nonumber
\end{align}
$\gamma'$s satisfy the following equations 
obtained by the maximizing the summand of the partition function 
with respect to the interaction parameters $\{\Gamma\}$:  
\[
\begin{split}
\frac{FG}{\gamma^{xy}_{bb} D}=\frac{1}{u}, 
\frac{GW}{\gamma^{yz}_{bb} D}=\frac{1}{u}, 
\frac{FW}{\gamma^{xz}_{bb} D}=\frac{1}{u}, \\
\frac{fg}{\gamma^{xy}_{hh} D}=\frac{1}{u}, 
\frac{gw}{\gamma^{yz}_{hh} D}=\frac{1}{u}, 
\frac{fw}{\gamma^{xz}_{hh} D}=\frac{1}{u}, \\
\frac{fG}{\gamma^{xy}_{hb} D}=\frac{1}{u}, 
\frac{gF}{\gamma^{yx}_{hb} D}=\frac{1}{u}, 
\frac{gW}{\gamma^{yz}_{hb} D}=\frac{1}{u}, \\
\frac{wG}{\gamma^{zy}_{hb} D}=\frac{1}{u}, 
\frac{fW}{\gamma^{xz}_{hb} D}=\frac{1}{u}, 
\frac{wF}{\gamma^{zx}_{hb} D}=\frac{1}{u}, \\
\frac{fgw}{\gamma^{xyz}_{hhh} D^2}=\frac{1}{v},
\frac{FGW}{\gamma^{xyz}_{bbb} D^2}=\frac{1}{v},
\frac{fgW}{\gamma^{xyz}_{hhb} D^2}=\frac{1}{v},
\frac{fwG}{\gamma^{xzy}_{hhb} D^2}=\frac{1}{v}, \\
\frac{gwF}{\gamma^{yzx}_{hhb} D^2}=\frac{1}{v},
\frac{wFG}{\gamma^{zxy}_{hbb} D^2}=\frac{1}{v},
\frac{gFW}{\gamma^{yxz}_{hbb} D^2}=\frac{1}{v},
\frac{fGW}{\gamma^{xyz}_{hbb} D^2}=\frac{1}{v}.
\end{split}
\]
Simplifying the above equations we obtain $\gamma^{ij}_{bb}=(k-1)^2 \gamma^{ij}_{hh}$, 
$\gamma^{ij}_{hb}=\gamma^{ji}_{hb}=(k-1)\gamma^{ij}_{hh}$, 
$\gamma^{ijk}_{bbb}=(k-1)^3 \gamma^{ijk}_{hhh}$, $\gamma^{ijk}_{hbb}=
(k-1)^2\gamma^{ijk}_{hhh}$ and $\gamma^{ijk}_{hhb}=(k-1)\gamma^{ijk}_{hhh}$. 
$F$, $G$, $W$, $f$, $g$, $w$, $D$ are simplified using the above relations.
Using these relations and setting $\rho_y=\rho_z$ and hence $\gamma^{xy}_{hh} 
=\gamma^{xz}_{hh}$, we reduce the number of independent equations 
for $\{\gamma\}$, given by,
\bea
f g = \frac{D \gamma^{xy}_{hh}}{u},
\label{eq:solve_gamma01}\\
g w = \frac{D \gamma^{yz}_{hh}}{u},
\label{eq:solve_gamma02}\\
f g w = \frac{\gamma^{xyz}_{hhh} D^2}{v}.
\label{eq:solve_gamma03}
\eea
We solve these three simultaneous equations to estimate the free energy. 
Maximizing the summand of the partition function with respect to $x_l$, 
$y_l$ and $z_l$ for $q=6$ and we obtain,
\bea
\frac{\left[\rho_x-\frac{\rho_x}{k}\right]^{k-1} \left[ 1-\rho+k^2
(\gamma^{xy}_{hh}+\gamma^{xz}_{hh}+\gamma^{yz}_{hh})+2 k^3 
\gamma^{xyz}_{hhh}\right]^k e^{\mu_1}}
{\left[1\!-\!\rho_x\!+\!\frac{\rho_x}{k}\right]^{k-1}\left[k-1\right]^{k-1}
\left[\frac{\rho_x}{k}-k\gamma^{xy}_{hh}-k\gamma^{xz}_{hh}-k^2\gamma^{xyz}_{hhh}\right]^k}&=&1,
\label{eq:rho_x}\\
\frac{\left[\rho_y-\frac{\rho_y}{k}\right]^{k-1} \left[ 1-\rho+k^2
(\gamma^{xy}_{hh}+\gamma^{xz}_{hh}+\gamma^{yz}_{hh})+2 k^3 
\gamma^{xyz}_{hhh}\right]^k e^{\mu_2}}
{\left[1\!-\!\rho_y\!+\!\frac{\rho_y}{k}\right]^{k-1}\left[k-1\right]^{k-1}
\left[\frac{\rho_y}{k}-k\gamma^{xy}_{hh}-k\gamma^{yz}_{hh}-k^2\gamma^{xyz}_{hhh}\right]^k}&=&1,
\label{eq:rho_y} \\
\frac{\left[\rho_z-\frac{\rho_z}{k}\right]^{k-1} \left[ 1-\rho+k^2
(\gamma^{xy}_{hh}+\gamma^{xz}_{hh}+\gamma^{yz}_{hh})+2 k^3 
\gamma^{xyz}_{hhh}\right]^k e^{\mu_3}}
{\left[1\!-\!\rho_z\!+\!\frac{\rho_z}{k}\right]^{k-1}\left[k-1\right]^{k-1}
\left[\frac{\rho_z}{k}-k\gamma^{xz}_{hh}-k\gamma^{yz}_{hh}-k^2\gamma^{xyz}_{hhh}\right]^k}&=&1.
\label{eq:rho_z}
\eea  
The expression of free energy for $q=6$, in terms of uniform, 
$N$-independent variables for a fixed $u$ and $v$ is given by, 
\bea
f(\rho_x,\rho_y,\rho_z, u)& =&  
-\frac{k-1}{k}\sum_{i=1}^{q/2} \rho_i \ln\rho_i 
-\sum_{i=1}^{q/2}\left[1-\frac{(k-1)\rho_i}{k} \right]
\ln\left[1-\frac{(k-1)\rho_i}{k} \right]
\nonumber \\ &&
+\sum_{i=1}^{q/2} \left[\rho_i - k^2 \sum_{j=1,j\neq i}\gamma^{ij}_{hh}-k^3\gamma_{hhh}\right]
 \ln \left[\rho_i - k^2 \sum_{j=1,j\neq i}^{q/2}\gamma^{ij}_{hh}-k^3\gamma_{hhh} \right]
\nonumber \\ &&
+\left[1-\rho+\frac{k^2}{2} \sum_{i,j=1,i\neq j}^{q/2}\gamma^{ij}_{hh}+2k^3\gamma_{hhh}\right] 
\nonumber \\ &&
\ln\left[1-\rho+\frac{k^2}{2}\sum_{i,j=1,i\neq j}^{q/2} \gamma^{ij}_{hh}+2k^3\gamma_{hhh}\right]
\nonumber \\ &&
-\frac{\rho}{k} \ln k
+\frac{k^2}{2} \sum_{i,j,i\neq j}^{q/2}\gamma^{ij}_{hh} \ln 
\left(\frac{k^2 \gamma^{ij}_{hh}}{u}\right)+
k^3 \gamma_{hhh}\ln \left(\frac{k^3\gamma_{hhh}}{v}\right).
\label{eq:free energy_q6}
\eea
For notational simplicity we have dropped the upper indices of $\gamma^{123}_{hhh}$.